\documentclass[reprint,floatfix,amsmath,amssymb,aps,prd,superscriptaddress,nofootinbib]{revtex4-2}

\usepackage{graphicx}
\usepackage{booktabs}
\usepackage{multirow}
\usepackage{siunitx}
\usepackage{bm}
\usepackage{microtype}
\usepackage[hidelinks]{hyperref}
\usepackage{xcolor}
\usepackage{enumitem}
\usepackage{orcidlink}

\graphicspath{{figures/}{.}}
\newcommand{\ttb}{t\bar t}
\newcommand{\che}{c_{\mathrm{hel}}}
\newcommand{\chanobs}{c_{\mathrm{han}}}
\newcommand{\mttbar}{m_{t\bar t}}
\newcommand{\gatt}{g_{A t\bar t}}

\begin{document}

\title{Distinguishing a pseudoscalar from a vector $\ttb$ resonance with top-quark spin correlations at the HL-LHC}
\makeatletter
\renewcommand{\andname}{,}
\makeatother
\author{A. Bellagroudi \orcidlink{0009-0002-7563-1176}}
\email{ahmed.bellagroudi@cern.ch}
\author{F. Fassi \orcidlink{0000-0002-6423-7213}}
\email{farida.fassi@cern.ch}
\affiliation{Laboratory of Condensed Matter and Interdisciplinary Sciences, Unit\'e de Recherche Labellis\'ee CNRST (URL-CNRST), Faculty of Sciences, Mohammed V University in Rabat, Rabat 1014, Morocco}

\begin{abstract}
The spin correlations imprinted on top-quark pairs provide a direct probe of the quantum numbers of a resonance decaying to $\ttb$. We compare a CP-odd type-II two-Higgs-doublet-model pseudoscalar $A$ with a spin-1 leptophobic topcolor $Z'$ in the dilepton channel. The two hypotheses differ in the parton-level helicity correlation by $\Delta\che\simeq0.43$, nearly independently of mass, and neutrino-weighted reconstruction preserves the separation with an effective dilution factor of about 0.66. At 400 and 800 GeV, the pseudoscalar normalization is anchored to the CMS HIG-22-013 coupling limits, while the $Z'$ hypothesis is rate-matched to the pseudoscalar so that the comparison primarily tests the resonance spin structure rather than an assumed rate difference. The leading-order $\ttb$-only statistical projection then gives expected separations of 2.27 and 2.74 standard deviations at $3000\,\mathrm{fb}^{-1}$. The reversed non-nested test gives the same Asimov separation to within numerical precision. Forcing the two selected signals to equal-yield, and hence removing the acceptance difference, gives pure spin-shape separations of 1.86 and 2.70 standard deviations. Profiling a Gaussian-constrained Standard Model $\che$-shape nuisance gives 2.07 and 2.63 standard deviations for an optimistic sideband-statistics benchmark; a weaker constraint of $\sigma_\theta=2\times10^{-3}$ reduces these values to 1.67 and 2.26. A local fixed-template signal-strength extrapolation, mapped through the adopted $g^4$ scaling, corresponds to five-standard-deviation discrimination at couplings approximately 22\% and 16\% above the reference values. The 1500 GeV point lies outside the CMS mass range and is shown separately as a $g=1$ projection. The reach is limited by the sub-percent signal fraction and by control of the mass-dependent Standard Model spin shape, rather than by the intrinsic resonance-spin separation. The remaining limitations include the leading-order normalization, fixed-width signal templates, the omission of non-$\ttb$ backgrounds
and signal--continuum interference, the simplified sideband-derived
mass-dependent nuisance model, and pseudoexperiment calibration.
\end{abstract}

\maketitle

\section{Introduction}
The top quark decays before hadronization, so the polarization information carried by the parent is transferred directly to the angular distributions of its decay products. In dileptonic $\ttb$ events the charged leptons have nearly maximal spin-analyzing power, making their correlated directions a particularly clean probe of the production spin-density matrix~\cite{Bernreuther2001,Bernreuther2015,ATLASspin}. The same density-matrix language has recently been used to formulate entanglement observables in top-quark pairs and to establish quantum entanglement experimentally near threshold~\cite{Afik2021,ATLASentanglement}. Related quantum-information studies continue to explore the structure and robustness of these correlations in Standard Model $gg$ and $q\bar q$ production~\cite{Jaloum2026,Bachain2026}.

Many extensions of the Standard Model predict neutral resonances above the $\ttb$ threshold. A two-Higgs-doublet model (2HDM) contains heavy CP-even and CP-odd neutral states, while topcolor and other dynamical electroweak-symmetry-breaking scenarios contain leptophobic vector bosons with enhanced couplings to the third generation~\cite{Branco2012,Carena2014,Hill1999,Harris2012}. Experimental searches usually identify these states through a localized excess or a peak--dip structure in the reconstructed $\mttbar$ spectrum. Such a search can establish a mass scale and constrain a production rate, but those quantities alone do not determine whether the state is scalar, pseudoscalar, or vector.
A spin assignment would therefore be among the first phenomenological questions after a discovery.

The resonance spin is imprinted on the helicity amplitudes of the produced top pair and is transmitted to the decay leptons. A CP-odd spin-zero state produces a predominantly spin-singlet $^1S_0$ configuration with a characteristic positive helicity correlation. A chiral spin-one state produces a different mixture of helicity configurations whose correlation depends on its vector and axial couplings. These ensemble-level differences are measurable even when the two signals have the same nominal mass and production rate.

Related work has addressed neighbouring questions. CP-even versus CP-odd discrimination within the spin-zero hypothesis has been studied in associated $Z\ttb$ production~\cite{Arco2025}, and CMS resonance searches use angular or spin-sensitive observables to improve sensitivity to heavy states in the $\ttb$ final state~\cite{CMSHIG,CMSB2G}. Spin and quantum-information observables have also been proposed as probes of the chiral structure of a $Z'$ in $\ttb$ production at future lepton colliders~\cite{ShivaSankar2026}, where the colourless, kinematically constrained initial state is advantageous. The present study instead asks a post-discovery question in inclusive dilepton $\ttb$: how well can reconstructed spin correlations distinguish a CP-odd resonance from a chiral spin-one resonance of the same mass, and what experimental conditions control that reach? To our knowledge, this specific $0^-$ versus $1^-$ comparison has not previously been formulated as an HL-LHC identification reach with explicit reconstruction dilution, equal-yield tests, bidirectional non-nested likelihoods, and a profiled dominant background spin nuisance.

We use a CP-odd type-II 2HDM pseudoscalar $A$ and the leptophobic topcolor Model IV $Z'$ as concrete benchmarks at masses of 400, 800, and 1500 GeV. Both signals are generated with matrix-element-level spin-preserving top decays, showered and passed through a fast detector simulation, and reconstructed with a neutrino-weighting algorithm. The principal observables are the helicity and transverse spin correlations in the Bernreuther basis. The headline likelihood uses a compact two-dimensional template in these observables, while one-dimensional, moment-based, and multivariate discriminants provide cross-checks.

The central result has two parts. At the level of the underlying physics, the two hypotheses are well separated: the parton-level helicity-correlation difference is $\Delta\che\simeq0.43$, is nearly independent of mass, and survives reconstruction with an effective dilution factor close to 0.66. At the level of the HL-LHC projection, however, the measurement is rate and background control limited. With the pseudoscalar normalized to the largest 400 and 800 GeV couplings not excluded by CMS HIG-22-013, and the $Z'$ rate-matched to it so that the test probes spin structure rather than an assumed rate difference, the leading-order $\ttb$-only statistical baseline reaches 2.27 and 2.74 standard deviations at $3000\,\mathrm{fb}^{-1}$. The 800 GeV result remains 2.70 after forcing equal selected signal yields and 2.63 after profiling an optimistic background-spin nuisance. Near threshold the Standard Model continuum is itself pseudoscalar-like, making the 400 GeV point intrinsically more fragile. At 1500 GeV the spin pattern remains distinct, but the benchmark is unreachable because only about twelve signal events survive selection.

The remainder of the paper is organized as follows. Section~II defines the two benchmark models, their spin-density structures, and the threshold qualification from toponium-like QCD dynamics. Section~III documents the simulation chain, decay treatment, normalization, and approximations. Section~IV describes the reconstruction, observables, and statistical procedure. Section~V presents the spin response, discriminant comparison, coupling and luminosity reach, robustness tests, and profiled background spin nuisance. Section~VI discusses the physical interpretation and experimental path forward, and Sec.~VII concludes.
\section{Theoretical framework and benchmark models}
\label{sec:models}
\subsection{CP-conserving type-II two-Higgs-doublet model}
The two-Higgs-doublet model extends the Standard Model scalar sector by two complex $SU(2)_L$ doublets, $\Phi_1$ and $\Phi_2$, with hypercharge $Y=1/2$. We assume CP conservation and a softly broken $Z_2$ symmetry, $\Phi_1\to\Phi_1$ and $\Phi_2\to-\Phi_2$, so that the scalar potential is~\cite{Branco2012}
\begin{align}
V(\Phi_1,\Phi_2)={}&m_{11}^2\Phi_1^\dagger\Phi_1+m_{22}^2\Phi_2^\dagger\Phi_2
-m_{12}^2\left(\Phi_1^\dagger\Phi_2+\mathrm{h.c.}\right) \notag\\
&+\frac{\lambda_1}{2}(\Phi_1^\dagger\Phi_1)^2
+\frac{\lambda_2}{2}(\Phi_2^\dagger\Phi_2)^2 \notag\\
&+\lambda_3(\Phi_1^\dagger\Phi_1)(\Phi_2^\dagger\Phi_2)
+\lambda_4(\Phi_1^\dagger\Phi_2)(\Phi_2^\dagger\Phi_1) \notag\\
&+\frac{\lambda_5}{2}\left[(\Phi_1^\dagger\Phi_2)^2+\mathrm{h.c.}\right],
\label{eq:2hdm_potential}
\end{align}
where all parameters are real. After electroweak symmetry breaking,
\begin{align}
\langle\Phi_i\rangle&=\frac{1}{\sqrt{2}}\begin{pmatrix}0\\v_i\end{pmatrix},\\
v&=\sqrt{v_1^2+v_2^2}=246~\mathrm{GeV},\\
\tan\beta&=\frac{v_2}{v_1}.
\end{align}
Three of the eight scalar degrees of freedom become the longitudinal components of the electroweak gauge bosons. The physical spectrum contains two CP-even neutral states $h$ and $H$, one CP-odd state $A$, and a charged pair $H^\pm$. Defining
\begin{equation}
M^2\equiv\frac{m_{12}^2}{\sin\beta\cos\beta},
\end{equation}
the pseudoscalar mass in the convention of Eq.~\eqref{eq:2hdm_potential} is
\begin{equation}
m_A^2=M^2-\lambda_5v^2.
\label{eq:mA}
\end{equation}
We work in the alignment limit, $\cos(\beta-\alpha)\to0$, in which the lighter CP-even state $h$ has Standard-Model-like couplings~\cite{Carena2014}. Alignment does not by itself guarantee that all heavy-scalar contributions to electroweak precision observables are negligible; those constraints also depend on the heavy-state masses and splittings. The present study therefore uses a phenomenological pseudoscalar benchmark motivated by the aligned type-II 2HDM rather than claiming a complete global fit of the scalar sector.

In a type-II Yukawa assignment, up-type fermions couple to $\Phi_2$, while down-type fermions and charged leptons couple to $\Phi_1$. The CP-odd interactions may be written as
\begin{align}
\mathcal L_{Aff}&=-i\,\frac{A}{v}\sum_f m_f\,\xi_A^f\,\bar f\gamma_5 f,\\
\xi_A^u&=\cot\beta,\qquad \xi_A^{d,\ell}=\tan\beta,
\label{eq:A_yukawa}
\end{align}
up to an overall convention-dependent sign. The top-quark term relevant here is
\begin{equation}
\mathcal L_{At\bar t}=-i\,\frac{m_t}{v}\cot\beta\,A\bar t\gamma_5t.
\label{eq:Att_full}
\end{equation}
In the normalization used below, $\gatt=\kappa_t^A=\cot\beta$. For $m_A>2m_t$, low or moderate $\tan\beta$, and a spectrum in which competing channels such as $A\to ZH$ are closed or suppressed, the top coupling controls both the loop-induced production $gg\to A$ and the decay $A\to\ttb$. The assumption on competing decays is important: the quoted $\sigma\times\mathrm{BR}$ values should not be interpreted as universal predictions over the full 2HDM parameter space.

The $i\gamma_5$ structure in Eq.~\eqref{eq:Att_full} is the essential input for the spin analysis. The CP-odd state produces the pair in a spin-singlet $^1S_0$ configuration. The resulting positive helicity correlation is a property of the decay spin state and is therefore largely independent of the production kinematics and resonance mass. The benchmark widths imported into the simulation are approximately $\Gamma_A=11.9$, 42.7, and 94.9 GeV, corresponding to $\Gamma_A/m_A=3.0\%$, $5.3\%$, and $6.3\%$ at 400, 800, and 1500 GeV, respectively.

\subsection{Leptophobic topcolor $Z'$}
Topcolor and related dynamical scenarios enlarge the gauge structure so that the third generation participates differently from the light fermions. A generic breaking pattern contains an additional neutral color-singlet vector boson associated with
\begin{equation}
U(1)_1\times U(1)_2\longrightarrow U(1)_Y.
\end{equation}
We use the leptophobic topcolor Model IV benchmark of Refs.~\cite{Hill1999,Harris2012}. The state couples to quarks, with enhanced relevance for the first and third generations, while charged-lepton couplings are absent. Its interaction can be written in the chiral form
\begin{equation}
\mathcal L_{Z'}=\frac{g_{Z'}}{2}Z'_\mu\,\bar q\gamma^\mu
\left[f_L(1-\gamma_5)+f_R(1+\gamma_5)\right]q,
\label{eq:Zp_full}
\end{equation}
which is equivalent to a $g_LP_L+g_RP_R$ parametrization after a redefinition of the coupling coefficients. The generated benchmark uses the one-chirality model-card choice denoted $f_1=1$, $f_2=0$. Because the notation for $f_{1,2}$ is not uniform across implementations, Eq.~\eqref{eq:Zp_full} defines the convention used in this manuscript and the numerical samples follow the validated UFO parameter card.

The $Z'$ is produced at tree level through $q\bar q$ annihilation and is studied through its $\ttb$ decay mode, which is enhanced by the third-generation coupling in the selected benchmark. Its total width scales quadratically with the overall coupling,
\begin{equation}
\Gamma_{Z'}(m_{Z'},g_{Z'})=g_{Z'}^2\,\widehat\Gamma(m_{Z'};f_L,f_R),
\label{eq:Zp_width}
\end{equation}
where $\widehat\Gamma$ contains the phase-space, spin, color, and flavour factors. The widths used in generation are taken directly from the run banners. In contrast to the pseudoscalar, the vector admits several helicity amplitudes and can generate both polarization and correlation structures. For the selected chiral Model IV benchmark, the mean helicity correlation is small and negative at parton-level. Other vector and axial assignments would produce different values, so the present result is a comparison of two specified models rather than a universal scalar-versus-vector theorem.

The two hypotheses also differ in production mode, polar-angle structure, and lineshape. In principle, $gg$ versus $q\bar q$ production and a vector decay angle could provide additional information. In an inclusive $pp$ analysis, however, the initial state direction is ambiguous, acceptance and reconstruction dilute angular asymmetries, and the mass lineshape is entangled with width and interference effects. We therefore design the baseline test to be dominated by the reconstructed top spin correlations rather than by the total rate or production angle information.

\subsection{Spin-density matrix and phenomenological handles}
The production and decay of the top pair can be expressed through the normalized two-spin density matrix
\begin{equation}
\rho =
\frac{1}{4}
\left[
\mathbf{1}_{2}\otimes\mathbf{1}_{2}
+ B_i^{+}\,\sigma_i\otimes\mathbf{1}_{2}
+ B_j^{-}\,\mathbf{1}_{2}\otimes\sigma_j
+ C_{ij}\,\sigma_i\otimes\sigma_j
\right].
\end{equation}
where $B_i^\pm$ describe the individual top and antitop polarizations and $C_{ij}$ is the spin correlation matrix. Charged leptons have nearly maximal spin analyzing power, so their directions in the parent top rest frames provide direct estimators of the relevant matrix elements~\cite{Bernreuther2001,Bernreuther2015}. The recent use of this matrix to formulate quantum information observables emphasizes that the same reconstructed angular data can test both Standard Model quantum correlations and the nature of a new production amplitude~\cite{Afik2021,ATLASentanglement,Jaloum2026,Bachain2026}.

For an $s$ channel spin zero resonance, rotational invariance removes any preferred polarization axis of the parent. The CP-odd decay vertex selects a spin-singlet state and produces a characteristic lepton opening angle distribution. A vector resonance admits a different correlation tensor whose form depends on the chiral couplings. These distinctions motivate the helicity and transverse observables defined in Sec.~\ref{sec:spinobs}. They also explain why the two-dimensional spin template captures most of the available information without requiring a highly flexible classifier.

\begin{table*}[t]
\caption{Principal theoretical and phenomenological handles distinguishing the benchmark hypotheses. The baseline statistical analysis is deliberately dominated by reconstructed spin correlation shapes rather than by the production mode or total rate.}
\label{tab:handles}
\begin{ruledtabular}
\begin{tabular}{lcc}
 & pseudoscalar $A$ & topcolor $Z'$ \\
\hline
spin/parity & $0^-$ & $1^-$ \\
underlying framework & type-II 2HDM, alignment limit & leptophobic topcolor Model IV \\
leading production & $gg$ through a top loop & $q\bar q$ at tree level \\
decay vertex & $i\gamma_5$ & chiral vector current \\
dominant $\ttb$ state & spin-singlet $^1S_0$ & coupling-dependent helicity mixture \\
helicity correlation & large and positive & small or negative for the benchmark \\
other potential handles & peak dip, no parent polarization & production angle, possible polarization \\
nominal lineshape & resonance plus continuum interference & Breit-Wigner like resonance \\
benchmark scope & CP-odd $A$, low/moderate $\tan\beta$ & one chirality $f_1=1,f_2=0$ choice \\
\end{tabular}
\end{ruledtabular}
\end{table*}

\subsection{Threshold dynamics and toponium}
Near the $\ttb$ threshold, slowly moving top pairs are affected by QCD Coulomb dynamics. The dominant color singlet pseudoscalar channel has the same $J^{PC}=0^{-+}$ and $^1S_0$ quantum numbers as the 2HDM pseudoscalar. The resulting toponium-like enhancement therefore carries a signal like helicity correlation~\cite{CMSToponium,Aguilar2024}. It is not a new elementary resonance of the type studied here, but it is an irreducible theoretical qualification of the 400 GeV benchmark. The Standard Model continuum near threshold is already closer to the pseudoscalar hypothesis than to the selected vector hypothesis, so the lowest mass point can be less discriminating despite its larger rate. This same physics underlies the observed threshold enhancement of top pair entanglement~\cite{ATLASentanglement} and is quantified through the mass dependence of the reconstructed background correlation in Sec.~\ref{sec:discussion}.

\section{Simulation and reconstruction}
\label{sec:simulation}
\subsection{Software chain and common settings}

Events are generated at $\sqrt{s}=14$~TeV with
\textsc{MadGraph5\_aMC@NLO} 3.5.5~\cite{MG5}, showered and
hadronized with \textsc{Pythia}~8~\cite{Pythia8}, and passed through
\textsc{Delphes}~3~\cite{Delphes} using the default CMS detector card. The
leading-order CTEQ6L1 parton distribution set is used
consistently at the matrix element and shower stages~\cite{Pumplin2002}.
Renormalization and factorization scales are evaluated event by event
using the default \textsc{MadGraph5\_aMC@NLO} dynamical scale prescription,
with \texttt{fixed\_ren\_scale=False},
\texttt{fixed\_fac\_scale=False},
\texttt{dynamical\_scale\_choice=-1}, and
\texttt{scalefact=1}. Generator-level cuts on leptons
and $b$ quarks are removed so that the event selection
is defined entirely after detector simulation.

The signal comprises six samples: $A$ and $Z'$ at 400, 800,
and 1500~GeV. The Standard Model $\ttb$ background is
generated as six statistically independent batches, giving
twelve generated samples in total. Every sample contains
$1.5\times10^{5}$ events. The six background batches therefore
provide $9.0\times10^{5}$ simulated events for template construction,
but they describe one physical process and one cross
section; the batch multiplicity is never used as a rate
factor. The analysis is restricted to $\ttb\to(b\ell^+\nu)(\bar b\ell^-\bar\nu)$
with $\ell=e,\mu$.

The topcolor $Z'$ was implemented in
\textsc{FeynRules}~\cite{Alloul2014b} following the Lagrangian of
Harris and Jain~\cite{Harris2012}. Separate right-handed up- and
down-type quark couplings were implemented in the model file to
realize the validated Model~IV parameter-card configuration,
denoted $f_1=1$, $f_2=0$ in the benchmark convention. The model was
exported in the UFO format~\cite{Degrande2012} for use in
\textsc{MadGraph5\_aMC@NLO} and validated by reproducing the
Harris--Jain parton-level cross section.

\subsection{Tree-level vector and Standard Model samples}
The topcolor $Z'$ and Standard Model background are tree-level processes and are generated with the top quark decay chains written directly into the matrix element. Schematically,
\begin{align}
q\bar q&\to Z'\to t\bar t,
& t&\to b\ell^+\nu,
& \bar t&\to\bar b\ell^-\bar\nu,\\
pp&\to t\bar t,
& t&\to b\ell^+\nu,
& \bar t&\to\bar b\ell^-\bar\nu.
\end{align}
The $Z'$ sample is summed over the light initial state flavours included in the validated model configuration and uses the chiral couplings defined in Sec.~\ref{sec:models}. Its width is taken from the run banner and follows the expected quadratic coupling scaling. The Standard Model background is normalized to the exact integrated generator weight from the banner,
\begin{equation}
\sigma_{\ttb\to e/\mu\,\mathrm{dilepton}}^{\rm LO}=25.1595~\mathrm{pb}.
\label{eq:bkg_xsec}
\end{equation}
Because the decay chains are present in the generated process, this value already includes the $e/\mu$ dilepton branching fraction. No additional factor of $(2/9)^2$ is applied. Combining the six batches changes only the Monte Carlo precision, not Eq.~\eqref{eq:bkg_xsec}.

\subsection{Pseudoscalar generation and the effective vertex}
The production process $gg\to A$ is loop induced. We generate it with the Higgs Characterization model~\cite{HCmodel}, in which the loop is represented by an effective $ggA$ interaction. The model restriction must retain both the effective gluon coupling and the top Yukawa. The CP-odd benchmark is selected with $\cos\alpha=0$, $\kappa_{Att}=\kappa_{Agg}=1$, and all CP-even couplings set to zero.\footnote{In the Higgs Characterization UFO, the generic neutral spin zero field $X_0$ carries PDG identifier 25 independently of its CP composition. The settings $\cos\alpha=0$, vanishing CP-even couplings, and nonzero CP-odd couplings define the state used here as purely CP odd; the PDG identifier is only a bookkeeping convention.}

The combined loop induced production and decay process exceeds the coupling order restriction of the model when written with the complete decay chain. We therefore generate $gg\to A$ through the effective vertex and decay the top pair with \textsc{MadSpin}~\cite{MadSpin}. The overall signal rate is normalized to the validated full 2HDM $\sigma\times\mathrm{BR}$ rather than to the effective vertex generation cross section. The nominal width fractions are 3.0\%, 5.3\%, and 6.3\% at 400, 800, and 1500 GeV.

\subsection{Treatment and validation of spin correlations}
The spin information studied here is present in the final state leptons only if the top decays are performed with knowledge of the production spin density matrix. In an initial validation sample, the hard process was terminated at stable top quarks and the decays were left entirely to the shower. The samples generated and reconstructed without errors, and their mass spectra and rates appeared reasonable, but the helicity correlation became nearly identical for the pseudoscalar, vector, and Standard Model hypotheses. This silent failure illustrates why rate level validation alone is insufficient for a spin analysis.

The production and decay matrix element is therefore retained explicitly for the tree level $Z'$ and background samples, while \textsc{MadSpin} is used for the loop induced pseudoscalar. The recovered Standard Model correlation and the stable $A$--$Z'$ separation provide internal validation that the spin information is preserved.

\subsection{Approximations and their scope}
Three approximations delimit the interpretation of the signal templates.

\emph{Effective pseudoscalar production.} The effective $ggA$ vertex omits the resolved finite-$m_t$ loop form factor that shapes the production lineshape. The decay spin state remains fixed by the CP-odd $At\bar t$ vertex, but the mass distribution and acceptance can change in a full loop calculation. The effective treatment is aligned with the benchmark template strategy used in the corresponding CMS search~\cite{CMSHIG}.

\emph{On-shell MadSpin decays.} The pseudoscalar top pair decay is performed in on-shell mode. The nominal width fractions, 3.0--6.3\%, are below the reconstructed $\mttbar$ resolution of order 20\%, so this approximation is expected to have a smaller effect on the reconstructed shape than the detector resolution. It nevertheless becomes relevant when the coupling and width are varied away from the generated benchmark.

\emph{Signal  continuum interference.} The coherent interference between $gg\to A\to\ttb$ and the QCD continuum is not included. Such interference produces a characteristic peak  dip structure and can redistribute tens of percent of the already small signal contribution within a fixed mass window~\cite{CarenaLiu,ATLASinterference}. Its leading impact is expected through the selected rate and mass-dependent spin density mixture. The quoted significances are therefore a pure resonance baseline, not a final interference-aware projection.

\subsection{Event selection and neutrino weighting}
Events are required to contain exactly two isolated, oppositely charged electrons or muons with $p_T>20$ GeV, at least two $b$-tagged jets, and missing transverse momentum above 30 GeV. We require $m_{\ell\ell}>20$ GeV to suppress low mass dilepton backgrounds and veto same flavour pairs within 15 GeV of the $Z$ mass. Tau leptons are not included because the additional neutrinos degrade the direct charged lepton spin analysis.

The two neutrinos make the dilepton kinematics underconstrained. Their six momentum components are constrained by the two $W$-mass shells, two top-mass shells, and the two measured components of missing transverse momentum, but the equations are nonlinear and can admit multiple or no real solutions. We use the neutrino-weighting method~\cite{D0NW}. The pseudorapidity of each neutrino is scanned over 21 equally spaced values in the interval $[-3,3]$, corresponding to a spacing of 0.3. For each $(\eta_{\nu_1},\eta_{\nu_2})$ point, the remaining kinematic equations are solved using $m_W=80.4$~GeV and $m_t=172.5$~GeV. Both assignments of the two leading $b$-tagged jets to the positively and negatively charged leptons are tested. For every kinematic solution, a weight is assigned according to the agreement between the vector sum of the reconstructed neutrino transverse momenta and the measured missing transverse momentum,
\begin{equation}
w =
\exp\!\left[
-\frac{
(\,p_x^{\nu_1}+p_x^{\nu_2}-E_x^{\rm miss}\,)^2+
(\,p_y^{\nu_1}+p_y^{\nu_2}-E_y^{\rm miss}\,)^2
}
{2\sigma_{\rm MET}^2}
\right],
\label{eq:nuweight}
\end{equation}
with $\sigma_{\rm MET}=15$~GeV. The solution with the largest weight, including the choice of $b$-jet--lepton assignment, defines the reconstructed top and antitop four-momenta. Events for which no real solution is found over the full scan are classified as reconstruction failures. Depending on the signal mass, the reconstruction succeeds for approximately 72--97\% of selected events.

The generated Standard Model normalization in Eq.~\eqref{eq:bkg_xsec} defines the baseline yields. It is not a complete HL-LHC background prediction. A complete experimental projection would require a consistent higher-order $\ttb$ normalization and explicit treatment of single-top, Drell--Yan plus heavy-flavour, diboson, $\ttb V$, and non-prompt-lepton backgrounds, together with their associated spin-sensitive shapes. Since the signal fraction is below one percent, even small additional components can matter if their angular distributions are aligned with either hypothesis.

\section{Spin observables and statistical procedure}
\label{sec:spinobs}
All momenta are first boosted to the $\ttb$ zero momentum frame. The reconstructed top direction defines the helicity axis $\hat{k}$. Each lepton is then boosted to the rest frame of its parent top, giving unit vectors $\hat b_+$ and $\hat b_-$. The principal observables are~\cite{Bernreuther2015,Arco2025}
\begin{align}
 \che &= \hat b_+\cdot\hat b_-, \\
 \chanobs &=2(\hat b_+\cdot\hat k)(\hat b_-\cdot\hat k)-\hat b_+\cdot\hat b_-.
 \label{eq:obs}
\end{align}
Both lie in $[-1,1]$. Additional observables from the complete $(\hat k,\hat r,\hat n)$ basis are used in the multivariate cross-check.

The statistical comparison uses 16 bins in the two-dimensional $(\che,\chanobs)$ plane, with four equal bins in each variable. The mass windows are 340--460, 680--920, and 1250--1750 GeV for the 400, 800, and 1500 GeV benchmarks. For expected bin counts $\nu_i^A=B_i+S_i^A$ and $\nu_i^{Z'}=B_i+S_i^{Z'}$, the directional Asimov statistic is
\begin{equation}
 q_{A\leftarrow Z'}=2\sum_i\left[
 \nu_i^{Z'}\ln\left(\frac{\nu_i^{Z'}}{\nu_i^A}\right)
 -\left(\nu_i^{Z'}-\nu_i^A\right)
 \right],
 \qquad Z=\sqrt{q},
 \label{eq:asimov}
\end{equation}
with $B+Z'$ treated as Asimov data and $B+A$ as the tested hypothesis~\cite{Cowan2011,pyhf}. The two signals are assigned the same preselection production normalization, while their selected acceptances are retained.

Because the hypotheses are non-nested, the reverse direction, $q_{Z'\leftarrow A}$, is evaluated independently. The two directions coincide to within numerical precision for the benchmark templates. We quote their minimum as the conservative identification significance. To isolate pure selected event shape information, the signal templates are also normalized to a common selected yield, taken as the average of the two physical selected yields. The resulting equal-yield test removes residual acceptance information while preserving the two-dimensional spin shapes.

The dominant simplified systematic is represented by a yield-preserving background tilt,
\begin{equation}
 B_i(\theta)=N_B\,
 \frac{b_i\left(1+\theta\,c_{\mathrm{hel},i}\right)}
 {\sum_j b_j\left(1+\theta\,c_{\mathrm{hel},j}\right)},
 \label{eq:tilt}
\end{equation}
where $b_i$ is the nominal normalized background template and $c_{\mathrm{hel},i}$ is the bin center value. The nuisance is Gaussian constrained, adding $\theta^2/\sigma_\theta^2$ to $-2\ln\mathcal L$, and is profiled separately under the two resonance hypotheses. The parameter calibration is such that $\sigma_\theta=8\times10^{-4}$ corresponds to approximately 0.3\% precision on the reconstructed background mean, the optimistic limit inferred from sideband counting statistics. This is a constrained one mode benchmark, not yet an explicit simultaneous sideband plus signal region fit.

\section{Results}
\subsection{Spin correlations and reconstruction response}
Table~\ref{tab:means} summarizes the mean helicity correlations. The pseudoscalar produces a large positive value, while the topcolor vector produces a small negative value at parton-level. The separation
\begin{equation}
 \Delta\che=\che(A)-\che(Z')\simeq0.43
\end{equation}
is almost independent of the resonance mass.

\begin{table}[t]
\caption{Mean helicity correlation at parton-level and after reconstruction. The maximum Monte Carlo statistical uncertainty is 0.006 at parton-level and 0.009 after reconstruction. The Standard Model entries are inclusive; the mass window dependence is shown in Fig.~\ref{fig:running}.}
\label{tab:means}
\begin{ruledtabular}
\begin{tabular}{lccc}
 & 400 GeV & 800 GeV & 1500 GeV \\
\hline
$A$ (parton) & $+0.334$ & $+0.336$ & $+0.344$ \\
$A$ (reco) & $+0.290$ & $+0.247$ & $+0.242$ \\
$Z'$ (parton) & $-0.084$ & $-0.095$ & $-0.082$ \\
$Z'$ (reco) & $+0.024$ & $-0.043$ & $-0.041$ \\
\hline
SM (parton) & \multicolumn{3}{c}{$+0.064$} \\
SM (reco) & \multicolumn{3}{c}{$+0.090$} \\
\end{tabular}
\end{ruledtabular}
\end{table}

Across the signal and inclusive background points, the reconstructed and parton-level means follow
\begin{equation}
 \che^{\mathrm{reco}}=0.656\,\che^{\mathrm{parton}}+0.039,
 \qquad R^2=0.963.
 \label{eq:dilution}
\end{equation}
The corresponding dilution of signal separation is 0.64, 0.67, and 0.66 at 400, 800, and 1500 GeV. The additive offset is common to the hypotheses and mostly cancels in their difference. In the idealized limit of the perfect reconstruction, the luminosity requirement associated with spin separation would improve by approximately $1/D^2\simeq2.2$.

\begin{figure*}[t]
 \centering
 \includegraphics[width=0.98\textwidth]{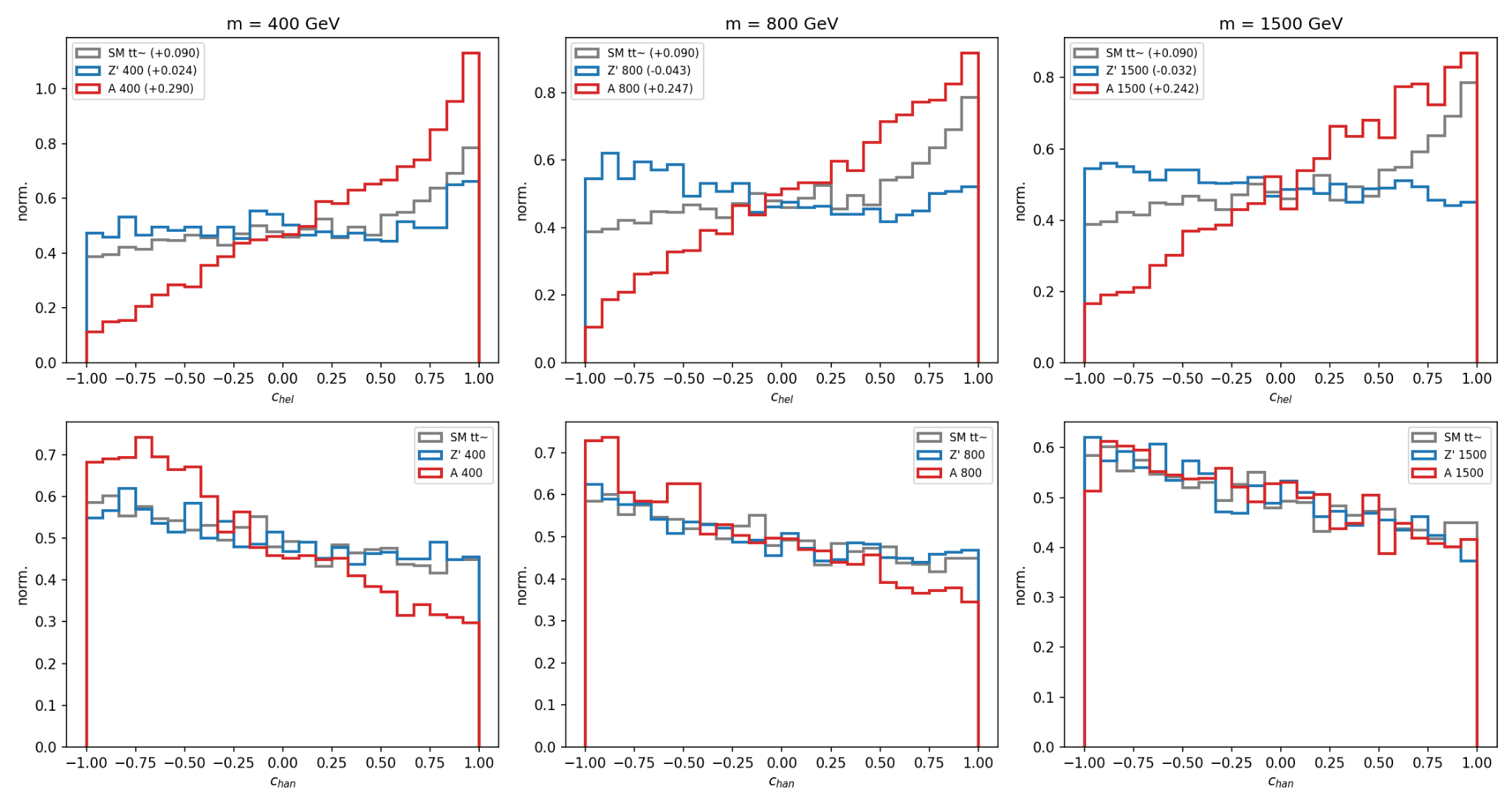}
 \caption{Reconstructed $\che$ (top) and $\chanobs$ (bottom) distributions for the pseudoscalar $A$, topcolor $Z'$, and Standard Model $\ttb$ samples. All curves are area normalized. The large difference in $\che$ persists over the full mass range, whereas $\chanobs$ adds information mainly at lower and intermediate masses.}
 \label{fig:shapes}
\end{figure*}
\begin{figure*}[t]
    \centering
    \includegraphics[width=0.98\textwidth]{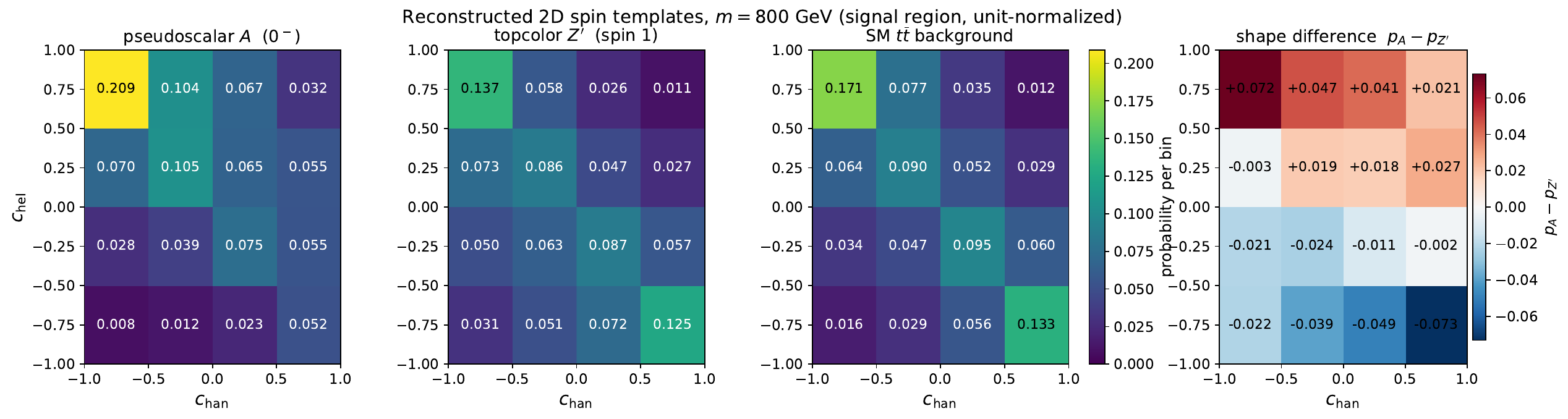}
    \caption{Reconstructed two-dimensional spin templates at
    $m=800$~GeV in the signal region. The first three panels show
    unit-normalized distributions in the $(\che,\chanobs)$ plane for
    the pseudoscalar $A$, the topcolor $Z'$, and the Standard Model
    $\ttb$ background. The final panel shows the normalized shape
    difference $p_A-p_{Z'}$. The $4\times4$ binning is identical to
    that used in the binned likelihood and therefore directly
    visualizes the spin-shape information entering the hypothesis
    discrimination.}
    \label{fig:spin2d}
\end{figure*}

The one-dimensional projections in Fig.~\ref{fig:shapes} make the
dominant helicity-correlation difference apparent, but the likelihood
uses the joint $(\che,\chanobs)$ distribution. Joint distributions of
these observables have previously been used to discriminate CP-even
and CP-odd spin-zero states in associated $Z\ttb$ production~\cite{Arco2025}.
Here we apply the same spin-correlation basis to the complementary
problem of distinguishing a direct pseudoscalar from a spin-one
$\ttb$ resonance. Figure~\ref{fig:spin2d} shows the reconstructed
two-dimensional templates for the strongest benchmark, $m=800$~GeV.

The pseudoscalar preferentially populates the region of positive
$\che$, whereas the vector hypothesis shifts probability toward
negative $\che$. The largest individual template difference occurs
for $\che\in[-1,-0.5]$ and $\chanobs\in[0.5,1]$, where
$p_A-p_{Z'}=-0.073$. The Standard Model template lies between the
two characteristic signal patterns.

\subsection{Discriminant comparison}
The pronounced ensemble-level differences in Fig.~\ref{fig:spin2d} do not imply event-by-event separation. As a multivariate cross-check, an XGBoost classifier~\cite{XGBoost} is trained independently at each resonance mass. The full-classifier AUC values are 0.650, 0.673, and 0.673 at 400, 800, and 1500~GeV, while the corresponding spin-only values are 0.628, 0.655, and 0.659. The modest improvement obtained after adding reconstructed kinematic variables indicates that most of the signal-hypothesis separation is already contained in the spin observables.

For the AUC study, the spin-only classifier uses the six reconstructed spin observables $(\che,\chanobs,c_r,c_n,\cos\theta_+,\cos\theta_-)$, while the full classifier additionally includes $\Delta\phi_{\ell\ell}$, $\cos\theta^\ast$, the transverse momenta of the two leptons, missing transverse momentum, and $\mttbar$, giving twelve input variables in total. The signal samples are divided into 70\% training and 30\% evaluation subsets using a fixed random seed. The classifier uses 300 trees with maximum depth four, learning rate 0.05, subsampling fraction 0.8, and column-subsampling fraction 0.8.

Table~\ref{tab:disc} compares four estimators at $3000\,\mathrm{fb}^{-1}$. The two-dimensional likelihood improves the mean-only result by 14\% at 800~GeV. The gain is real but modest, and the BDT does not outperform the compact spin-template likelihood. The BDT significance quoted in Table~\ref{tab:disc} is a separate cross-check: the available Standard Model background samples contain only $\che$ and $\chanobs$, so this classifier is restricted to the same two reconstructed spin variables used in the binned likelihood. It should therefore be interpreted as a nonlinear consistency check of the two-dimensional template result rather than as an independent higher-dimensional multivariate discriminant.

\begin{table}[t]
\caption{Expected baseline separation at $3000\,\mathrm{fb}^{-1}$ for different discriminants.}
\label{tab:disc}
\begin{ruledtabular}
\begin{tabular}{lcccc}
$m$ & mean & 1D $\che$ & 2D $(\che,\chanobs)$ & BDT \\
\hline
400 GeV & 1.92 & 2.31 & 2.27 & 2.21 \\
800 GeV & 2.40 & 2.45 & 2.74 & 2.46 \\
1500 GeV & 0.03 & 0.03 & 0.05 & 0.15 \\
\end{tabular}
\end{ruledtabular}
\end{table}

\begin{figure*}[t]
 \centering
 \includegraphics[width=0.98\textwidth]{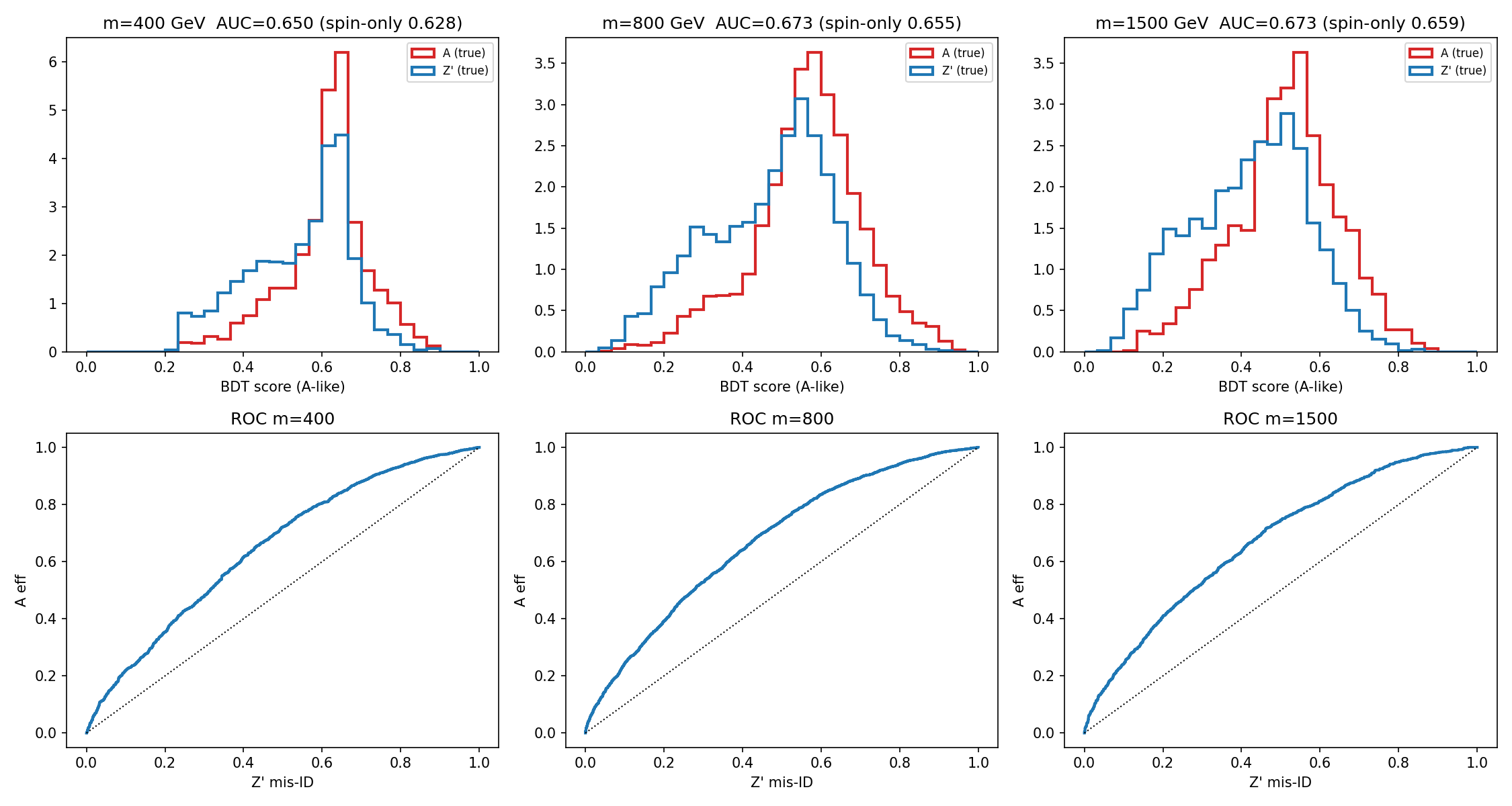}
 \caption{XGBoost score distributions and ROC curves for the $A$--$Z'$ classification. The AUC values are moderate, consistent with substantial event-level overlap. The modest gain from adding reconstructed kinematic variables to the six spin observables supports the interpretation that the discrimination is primarily driven by spin information.}
 \label{fig:bdt}
\end{figure*}

\subsection{Rate limitation, robustness, and coupling reach}
\label{sec:reach}
\begin{figure}[t]
 \centering
 \includegraphics[width=\columnwidth]{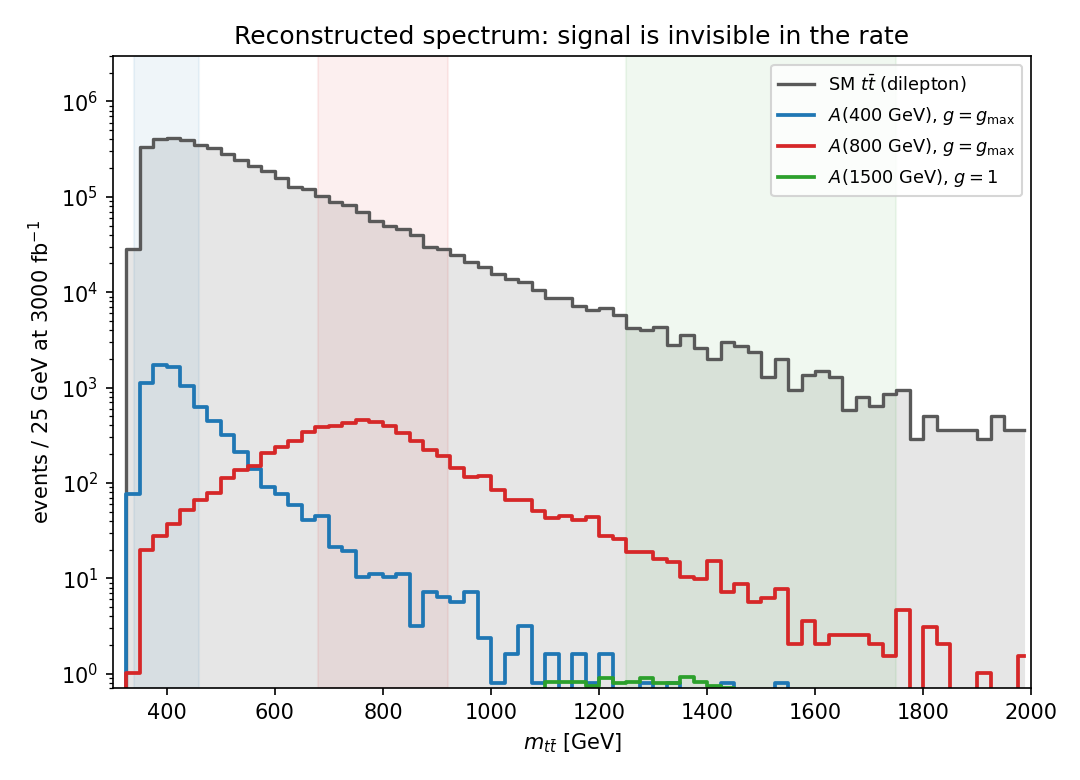}
 \caption{Expected reconstructed $\mttbar$ yields at $3000\,\mathrm{fb}^{-1}$ in the LO $\ttb$-only baseline. The 400 and 800 GeV signals use the reference couplings obtained from CMS HIG-22-013~\cite{CMSHIG}; the 1500 GeV curve is a $g=1$ projection outside that search range. Shaded bands mark the analysis windows.}
 \label{fig:spectrum}
\end{figure}

Figure~\ref{fig:spectrum} shows the reconstructed spectrum in baseline normalization. The signal is nowhere visible as a conventional rate excess. The expected yields at $3000\,\mathrm{fb}^{-1}$ are approximately $(N_S,N_B)=(5849,1.72\times10^6)$, $(3411,5.63\times10^5)$, and $(12,4.30\times10^4)$ at 400, 800, and 1500 GeV, corresponding to $S/B=0.34\%$, $0.61\%$, and $0.03\%$.

\begin{table}[t]
\caption{Physical normalization and pure shape robustness tests at $3000\,\mathrm{fb}^{-1}$. The conservative value is the minimum of the two non-nested directions.}
\label{tab:robustness}
\begin{ruledtabular}
\begin{tabular}{lcccc}
$m$ & $Z_{A\leftarrow Z'}$ & $Z_{Z'\leftarrow A}$ & conservative & equal-yield \\
\hline
400 GeV & 2.27 & 2.27 & 2.27 & 1.86 \\
800 GeV & 2.74 & 2.74 & 2.74 & 2.70 \\
1500 GeV & 0.05 & 0.05 & 0.05 & 0.06 \\
\end{tabular}
\end{ruledtabular}
\end{table}

At 800 GeV, forcing equal selected yields changes the separation only from 2.74 to 2.70 standard deviations, demonstrating that the result is almost entirely driven by the reconstructed spin shape. At 400 GeV, the reduction from 2.27 to 1.86 shows that acceptance differences supplement the weaker near threshold shape separation. The reverse hypothesis direction does not weaken either benchmark.

\begin{figure}[t]
 \centering
 \includegraphics[width=\columnwidth]{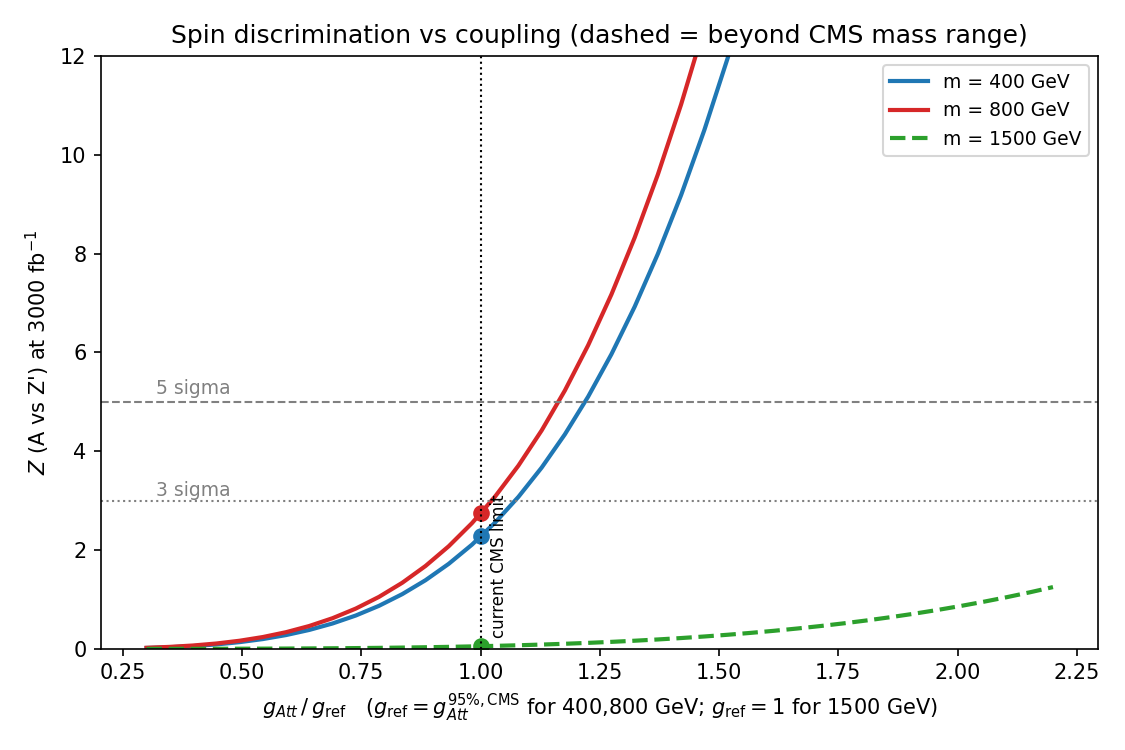}
 \caption{Statistical baseline separation as a function of the coupling at $3000\,\mathrm{fb}^{-1}$. For 400 and 800 GeV, $g_{\rm ref}$ is the CMS 95\% limit; for the 1500 GeV projection $g_{\rm ref}=1$. The templates are held fixed, so the 5$\sigma$ crossings are local extrapolations.}
 \label{fig:coupling}
\end{figure}

The fixed-template coupling scan is shown in Fig.~\ref{fig:coupling}. At the reference couplings $g_{\rm ref}=0.527$ and 1.19, the 400 and 800 GeV benchmarks reach 2.27 and 2.74 standard deviations. Five standard deviation discrimination occurs at $g/g_{\rm ref}=1.22$ and 1.16. These values are local diagnostics rather than self-consistent exclusion reinterpretations because the signal widths and the width-dependent CMS limits change with the coupling.

\begin{figure}[t]
 \centering
 \includegraphics[width=\columnwidth]{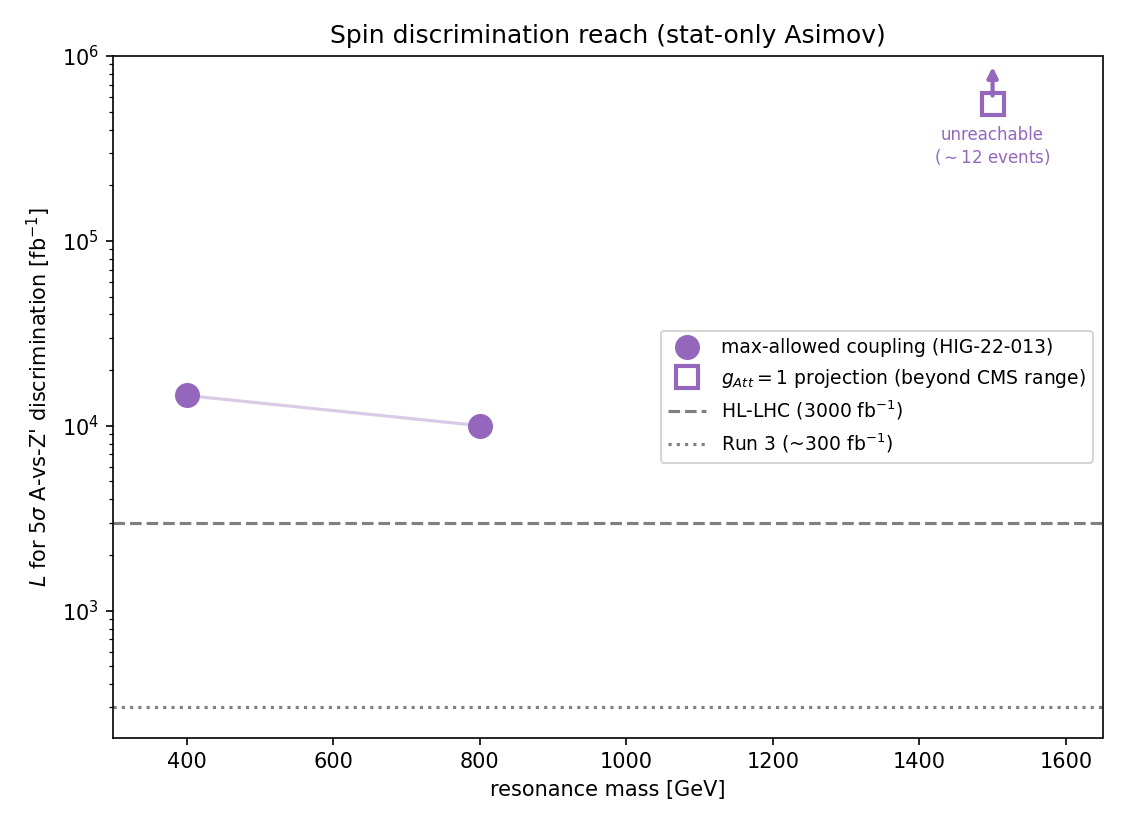}
 \caption{Luminosity required for a 5$\sigma$ $A$--$Z'$ spin discrimination. Filled points use the CMS-normalized 400 and 800 GeV benchmarks. The 1500 GeV $g=1$ projection is far off scale and is indicated by an open marker and upward arrow.}
 \label{fig:luminosity}
\end{figure}

Figure~\ref{fig:luminosity} gives the equivalent luminosity reach. The 400 and 800 GeV points require approximately $1.46\times10^4$ and $1.00\times10^4\,\mathrm{fb}^{-1}$ in the statistical baseline. The 1500 GeV projection has $Z\simeq0.05$ at the HL-LHC, corresponding to a naive scaling of order $3\times10^7\,\mathrm{fb}^{-1}$ and is shown as off scale.

\subsection{Background spin shape control}
\begin{figure}[t]
 \centering
 \includegraphics[width=\columnwidth]{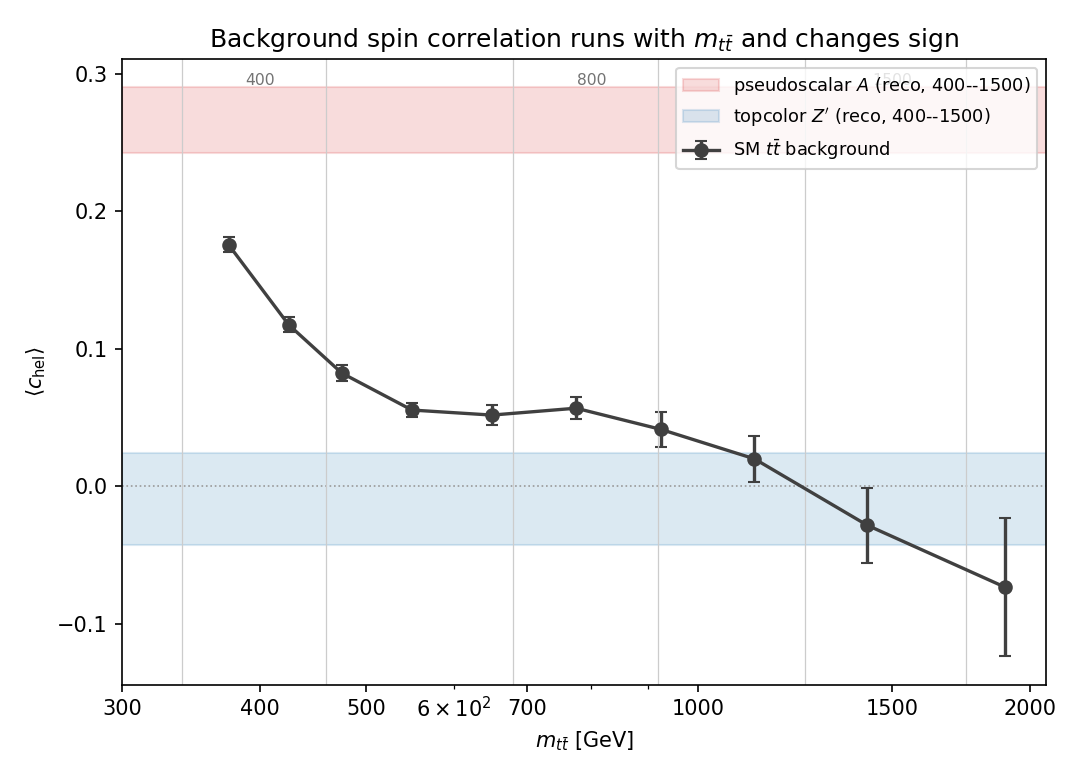}
 \caption{Mass dependence of the simulated reconstructed Standard Model helicity correlation, compared with the ranges spanned by the pseudoscalar and topcolor-vector benchmarks. The benchmark signal windows are indicated at the top.}
 \label{fig:running}
\end{figure}

The simulated reconstructed Standard Model helicity correlation is not constant with $\mttbar$. Figure~\ref{fig:running} shows that it decreases from approximately $+0.18$ near threshold, crosses zero near 1.2 TeV, and becomes negative at high mass. This behaviour is consistent with the experimentally established threshold spin-singlet enhancement~\cite{ATLASentanglement}. It explains why the 400 GeV benchmark is harder than 800 GeV despite its larger signal yield: the threshold background is already closer to the pseudoscalar state.

\begin{figure}[t]
 \centering
 \includegraphics[width=\columnwidth]{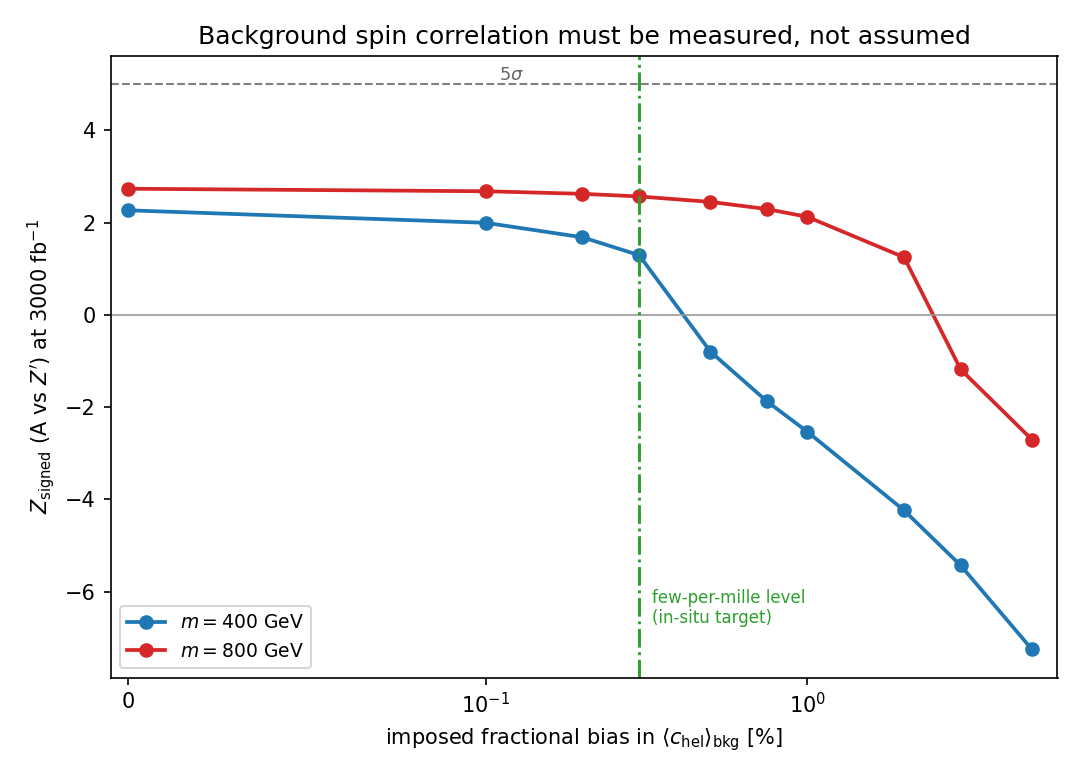}
 \caption{Signed Asimov preference versus an imposed fractional bias in the background mean helicity correlation. This is an unprofiled mismodelling test, not a conventional nuisance parameter uncertainty.}
 \label{fig:bias}
\end{figure}

As a robustness test, the background is first distorted by an imposed yield-preserving shift of its mean helicity correlation and tested against nominal templates. Figure~\ref{fig:bias} is deliberately signed: negative values mean that the unmodelled bias makes the incorrect resonance hypothesis preferred. At 400 GeV, a 0.5\% bias already reverses the preference, while the 800 GeV result remains positive until biases between 1\% and 3\%.

\begin{figure}[t]
 \centering
 \includegraphics[width=\columnwidth]{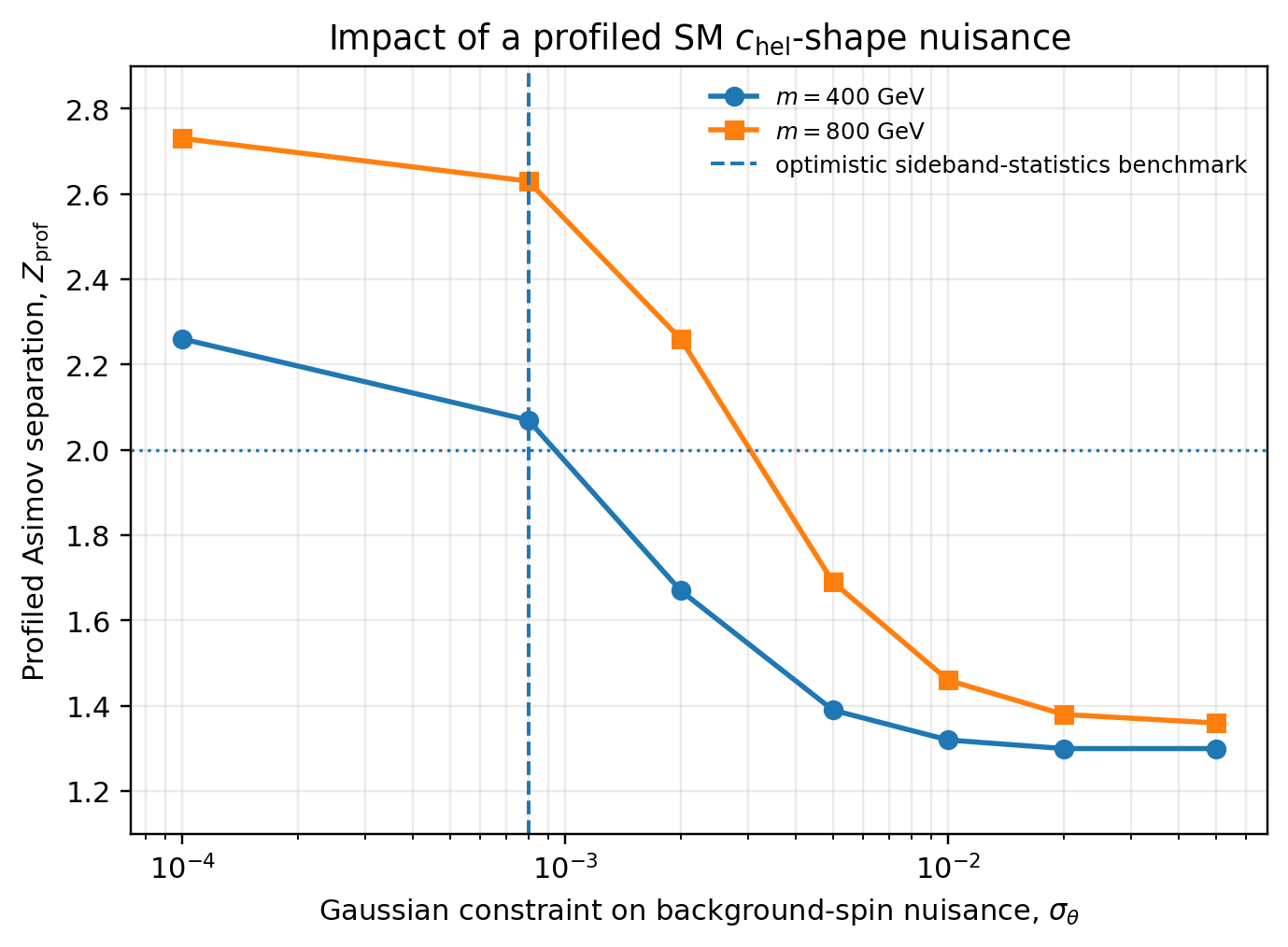}
 \caption{Profiled $A$--$Z'$ Asimov separation versus the Gaussian constraint on the yield-preserving Standard Model $\che$-shape nuisance. The vertical line marks the optimistic sideband statistics benchmark.}
 \label{fig:profiled}
\end{figure}

The same deformation is then introduced as the Gaussian constrained nuisance of Eq.~\eqref{eq:tilt} and profiled. The scan in Fig.~\ref{fig:profiled} and Table~\ref{tab:profiled} shows that the optimistic sideband statistics constraint, $\sigma_\theta=8\times10^{-4}$, retains most of the baseline sensitivity: $2.27\to2.07$ at 400 GeV and $2.74\to2.63$ at 800 GeV. A weaker constraint of $2\times10^{-3}$ gives 1.67 and 2.26. For very weak constraints the significance approaches approximately 1.30 and 1.38, showing that the single $\che$ tilt cannot absorb all of the two-dimensional spin information.

\begin{table}[t]
\caption{Profiled Asimov separation for representative Gaussian constraints on the dominant SM $\che$-shape nuisance.}
\label{tab:profiled}
\begin{ruledtabular}
\begin{tabular}{lcc}
$\sigma_\theta$ & 400 GeV & 800 GeV \\
\hline
$1\times10^{-4}$ & 2.26 & 2.73 \\
$8\times10^{-4}$ & 2.07 & 2.63 \\
$2\times10^{-3}$ & 1.67 & 2.26 \\
$5\times10^{-3}$ & 1.39 & 1.69 \\
$2\times10^{-2}$ & 1.30 & 1.38 \\
\end{tabular}
\end{ruledtabular}
\end{table}

The profiled scan is more realistic than the unprofiled bias study, but it does not yet determine the nuisance from data like sideband channels. Because $\langle\che\rangle_{\rm bkg}$ runs strongly with $\mttbar$, the decisive experimental step is a simultaneous mass-dependent fit of low sideband, signal region, and high sideband. The present profile should therefore be read as a controlled bracket: it quantifies how the reach depends on the constraint, while leaving the signal region interpolation uncertainty to a dedicated implementation.

\section{Discussion}
\label{sec:discussion}
The central result separates intrinsic spin information from experimental reach. The parton-level difference $\Delta\che\simeq0.43$ is large and nearly mass independent, and approximately two thirds survives reconstruction. The deterioration of the reach with increasing mass is therefore not caused by a loss of spin separation. It is caused by the falling selected signal yield. At 1500 GeV the reconstructed $A$ and $Z'$ spin distributions remain distinct, but the $g=1$ benchmark contributes only about twelve selected events at $3000\,\mathrm{fb}^{-1}$ and is consequently unreachable.

The 800 GeV benchmark provides the cleanest demonstration of genuine spin-shape discrimination. Its equal selected yield result, 2.70 standard deviations, is almost identical to the physical normalization value of 2.74. The full two-dimensional likelihood also improves on the mean-only estimator, while the BDT fails to exceed the compact spin template. These checks show that the sensitivity is carried by structured ensemble-level spin information rather than by an acceptance artifact, a residual mass feature, or classifier flexibility.

The 400 GeV point illustrates a qualitatively different limitation. Near threshold, Standard Model $\ttb$ production contains a large spin-singlet component. The reconstructed background correlation in the 400 GeV window is therefore already close to the pseudoscalar value. Equalizing the selected signal yields reduces the significance from 2.27 to 1.86, and the profiled background spin nuisance has a larger impact than at 800 GeV. The same threshold dynamics that enhances entanglement and motivates toponium studies thus works against a post-discovery $A$--$Z'$ identification~\cite{ATLASentanglement,CMSToponium,Aguilar2024}.

The background running figure makes this limitation explicit. The simulated reconstructed $\langle\che\rangle_{\rm bkg}$ decreases from about $+0.18$ near threshold, crosses zero near 1.2 TeV, and reaches negative values at high mass. A single inclusive background spin number is therefore not transferable between the three signal windows. The unprofiled bias scan shows how quickly an unmodelled distortion can reverse the signed preference, especially at 400 GeV. The constrained profile gives a more physical interpretation: for the optimistic sideband statistics benchmark, the significances remain 2.07 and 2.63, while a $2\times10^{-3}$ constraint gives 1.67 and 2.26. The nonzero weak constraint floor indicates that the two-dimensional signal difference is not completely degenerate with one linear $\che$ tilt.

The current nuisance model is nevertheless only a bracket. The nominal $\sigma_\theta=8\times10^{-4}$ value is motivated by the counting power of roughly five million sideband events, but the decisive uncertainty is the interpolation of the strongly mass-dependent spin shape into the signal window. A realistic measurement should fit low sideband, signal region, and high sideband simultaneously, with at least normalization, slope, and curvature modes in $\mttbar$. Detector response and shower variations should then be represented by additional correlated shape templates rather than absorbed into a single phenomenological tilt.

Several conventional uncertainties have different implications for this non-nested comparison. A fully common rate variation of the shared background or of two equally normalized signal hypotheses largely cancels in the preference between $A$ and $Z'$. By contrast, a shape change aligned with the difference of the two spin templates can be strongly amplified by the sub-percent signal fraction. The most relevant missing ingredients are therefore not only the inclusive $\ttb$ scale and PDF uncertainty, but also shower and hadronization variations, finite Monte Carlo statistics in the two-dimensional bins, non-$\ttb$ backgrounds with nontrivial angular structure, and the coupling-dependent interference and width effects.

The coupling scan should be interpreted accordingly. The scan is fundamentally a common signal-strength rescaling of the two rate-matched hypotheses, $\mu=N_S/N_{S,\mathrm{ref}}$. Within the local fixed-template benchmark this signal-strength change may be mapped approximately onto a coupling rescaling, $g/g_{\mathrm{ref}}\simeq\mu^{1/4}$, using the adopted $g^4$ dependence. The resulting 5$\sigma$ crossings, $g/g_{\mathrm{ref}}=1.22$ and 1.16 at 400 and 800 GeV, should therefore be interpreted only as indicative local diagnostics. They are not self-consistent new exclusion limits: the width scales approximately as $g^2$, the mass window acceptance changes, and the CMS limit itself is width-dependent. Regenerated or morphed width-dependent templates and coherent interference are needed before those crossings can be interpreted quantitatively.

The event selection also offers limited room for a simple gain. Relaxing from two $b$ tags to one increases the selected signal by a factor 1.67, but the background rises by 2.04 and the spin separation changes by only a factor 0.948. The resulting significance gain is about 1.11, while additional non-$\ttb$ backgrounds become more important. This does not justify replacing the two tag baseline. A more promising extension is the lepton+jets channel, whose larger branching fraction could provide roughly six times more signal events. Its gain is governed by the effective analyzing power of the hadronic side, combinatorial reconstruction, and larger backgrounds; it is therefore a substantial follow-up analysis rather than a trivial rescaling.

The remaining work before a final experimental projection is clear: a consistent higher-order normalization and scale/PDF treatment; explicit non-$\ttb$ templates; parton shower and detector variations; interference-aware pseudoscalar generation; width-dependent signal morphing; a simultaneous mass-dependent sideband fit; and pseudoexperiment calibration of the non-nested test statistic.

\section{Conclusions}
Dilepton top quark spin correlations produce clearly distinct ensemble-level distributions for the CP-odd pseudoscalar and chiral topcolor-vector benchmarks. The parton-level helicity correlation difference is approximately 0.43, nearly independent of mass, and survives neutrino weighted reconstruction with an effective dilution of about 0.66.

At the CMS normalized 400 and 800 GeV reference points, the LO $\ttb$-only statistical baseline gives 2.27 and 2.74 standard deviations at $3000\,\mathrm{fb}^{-1}$. The reverse non-nested hypothesis direction gives the same result. Removing acceptance differences gives pure spin shape separations of 1.86 and 2.70, establishing that the 800 GeV sensitivity is almost entirely shape driven. Profiling the dominant simplified Standard Model $\che$-shape nuisance gives 2.07 and 2.63 for an optimistic sideband statistics constraint, and 1.67 and 2.26 for a weaker $2\times10^{-3}$ constraint.

The observable therefore works, but the measurement is rate and background control limited. The 1500 GeV benchmark is unreachable at the chosen $g=1$ projection, while the 400 GeV point is intrinsically difficult because the threshold background is pseudoscalar like. A local fixed-template scan places 5$\sigma$ separation 16--22\% above the present 400 and 800 GeV reference couplings, but a self-consistent interpretation requires width-dependent and interference-aware templates.

A more realistic extension is therefore not primarily a more complex
classifier, but a simultaneous mass-dependent sideband and signal-region
fit that determines the Standard Model spin shape in situ. Subject to the
stated normalization, background, interference, and calibration limitations,
the study demonstrates the potential of spin correlations for post-discovery
resonance identification and quantifies the level of spin-shape control
required to exploit this information.

\begin{acknowledgments}
The work of A.~Bellagroudi is funded by the National Center for Scientific and Technical Research (CNRST) under the PhD-Associate Scholarship (PASS). 

\end{acknowledgments}

\section*{Data Availability}
The custom \textsc{FeynRules} model file and derived UFO implementation for the topcolor $Z'$, together with the event-generation cards and analysis code used in this study, are available from the corresponding author upon reasonable request. The Monte Carlo samples can be regenerated from the documented generator configuration and parameter cards.

\end{document}